# Chemical Bonding Governs Complex Magnetism in MnPt$_5$P

*Xin Gui,[1] Ryan A. Klein,[2,3] Craig M. Brown,[3] Weiwei Xie[1]\**

1. Department of Chemistry, Louisiana State University, Baton Rouge, LA, 70803, USA
2. Chemistry and Nanoscience Department, National Renewable Energy Laboratory, Golden, CO, 80401, USA
3. NIST Center for Neutron Research, National Institute of Standards and Technology, Gaithersburg, MD, 20899, USA

## *ABSTRACT*

Subtle changes in chemical bonds may result in dramatic revolutions in magnetic properties in solid state materials. MnPt$_5$P, a new derivative of the rare-earth-free ferromagnetic MnPt$_5$As, was discovered and is presented in this work. MnPt$_5$P was synthesized and its crystal structure and chemical composition were characterized by X-ray diffraction as well as energy-dispersive X-ray spectroscopy. Accordingly, MnPt$_5$P crystallizes in the layered tetragonal structure with the space group *P*4/*mmm* (No. 123), in which the face-shared Mn@Pt$_{12}$ polyhedral layers are separated by P layers. In contrast to the ferromagnetism observed in MnPt$_5$As, the magnetic properties measurements on MnPt$_5$P show antiferromagnetic ordering occurs at ~188 K with a strong magnetic anisotropy in and out of the *ab*-plane. Moreover, a spin-flop transition appears when a high magnetic field is applied. An A-type antiferromagnetic structure was obtained from the analysis of powder neutron diffraction (PND) patterns collected at 150 K and 9 K. Calculated electronic structures imply that hybridization of Mn-3*d* and Pt-5*d* orbitals are critical for both the structural stability and observed magnetic properties. Semi-empirical molecular orbitals calculations on both MnPt$_5$P and MnPt$_5$As indicate that the lack of 4*p* character on the P atoms at the highest occupied molecular orbital (HOMO) in MnPt$_5$P may cause the different magnetic behavior in MnPt$_5$P compared to MnPt$_5$As. The discovery of MnPt$_5$P, along with our previously reported MnPt$_5$As, parametrizes the end points of a tunable system to study the chemical bonding which tunes the magnetic ordering from ferromagnetism to antiferromagnetism with strong spin-orbit coupling (SOC) effect.

## *Introduction*

Chemical bonding concepts are critical for understanding and predicting the chemical compositions, structural stabilities, and resulting exotic physical properties in solid state materials.[1-5] Generally, chemical bonds in solids are directly related to the total energy of formation, which primarily originates from ionic and covalent interactions. Thus, compounds with the same structure show similar energies of formations with distinct chemical bonds due to slight differences in chemical compositions and atomic distances. The subtle changes in chemical bonds become decisive for the physical behaviors, for example, magnetism.[6] Over the past decades, complex intermetallic compounds have been synthesized containing magnetically active Mn atoms with various Mn-Mn configurations, allowing the studies of magnetic exchange as a function of atomic distances and chemical bonding interactions.[7-14] The previous studies revealed a consistent conclusion that magnetic ordering originated from Mn-Mn interlayer exchange interaction is extremely sensitive to lattice constants of the unit cell and atomic distances due to thermal expansion.[15-18] Moreover, the transformation between ferromagnetic (FM) states and antiferromagnetic (AFM) states can be controlled and manipulated by temperature, pressure, magnetic field (spin-flop), and chemical doping in several systems, such as $RMn_6X_6$[7-9] and $RMn_2X_2$ (R = rare-earth elements; X = Sn/Ge)[10-14]. Such FM-AFM transitions are suitable for heat assisted magnetic recording (HAMR) technologies.[19-22]

$MnPt_5As$ crystallizes in a layered tetragonal structure with the space group of *P*4/*mmm*, analogous to one of the well-known heavy fermion superconductors, $CeCoIn_5$.[23,24] $MnPt_5As$ was reported to order ferromagnetically with a large magnetic moment of 3.6 $\mu_B$ on Mn at room temperature ($T_C$ ~ 301 K), indicating a strong Mn-Mn FM exchange interaction.[25] The theoretical assessment on $MnPt_5As$ showed the Mn-Mn antibonding interaction dominates the density of states above and below the Fermi energy allowing the structure to relax through spin polarization. Moreover, the Mn-Mn distance along the *c*-axis is nearly twice as long as the distance in the *ab*-plane, which induces a large anisotropy in and out of the *ab*-plane. A Goodenough-Kanamori style analysis of the structure suggests both intralayer FM superexchange interaction along the 90° Mn–Pt–Mn pathway and a much weaker interlayer FM superexchange interaction along the Mn–Pt–As–Pt–Mn pathway, leading to long-range FM ordering.

To further understand how chemical bonding rules the magnetic, the As ions in MnPt$_5$As were substituted by the isovalent but much smaller P ions. A new compound, MnPt$_5$P, was designed and synthesized. MnPt$_5$P crystallizing in the same structure as MnPt$_5$As and was determined to order antiferromagnetically with T$_N$ ~ 188 K. Electronic structure and molecular orbital calculations indicate that the shorter Mn-Mn distance due to the lack of Mn-3$d$ and P-4$p$ orbital hybridization in MnPt$_5$P is crucial for the antiferromagnetic interactions. The new antiferromagnetic MnPt$_5$P, along with the previously reported MnPt$_5$As, can be an ideally tunable system to investigate the impact of chemical bonding on magnetic ordering.

*Experimental Section*

**Sample preparation:** The procedures for the synthesis of MnPt$_5$As (ref. 25) were adapted for the synthesis of polycrystalline MnPt$_5$P. Mn powder, Pt powder and red P powder were evenly mixed with a molecular ratio of Mn: Pt: P = 1:5:1. The mixture was ground and pressed into a pellet, and then placed into an alumina crucible. The crucible was sealed into an evacuated silica tube (<10$^{-5}$ torr) and then heated to 1050 °C at a rate of 30 °C per hour. The sample was slowly cooled to room temperature for two weeks after annealing at 1050 °C for two days. Small single crystals (~0.8 × 0.8 × 0.2 mm$^3$) were obtained from the product chunk. Structural characterization and physical properties measurements were performed on the single crystals obtained. MnPt$_5$P is stable to both air and moisture.

**Phase Identification:** The powder X-ray diffraction (PXRD) patterns of the synthesized sample were measured with a Rigaku MiniFlex[†] 600 powder X-ray diffractometer equipped with a Cu K$_\alpha$ radiation ($\lambda$ = 1.5406 Å, Ge monochromator). Data were collected over scattering angle, 2$\theta$, from 5° to 90° with a step of 0.005° at a rate of 0.1°/min. Rietveld analysis was performed using the Fullprof Suite to obtain the weight percentage of obtained phases.[26]

**Structure Determination:** Single crystal X-ray diffraction experiments were conducted on a Bruker Apex II diffractometer equipped with Mo radiation ($\lambda_{K\alpha}$ = 0.71073 Å) at room temperature to determine the crystal structure of MnPt$_5$P. Data for multiple crystallites (~10× 60× 60 µm$^3$) from different batches were collected to ensure homogeneity. Crystals were mounted on a Kapton loop and protected by glycerol. Four distinct combinations of positions for crystals and detector were determined by the software according to the pre-determined unit cell. The scanning 2θ width was set to 0.5° with the exposure time of 10 s. The crystal structure was solved based on direct methods and full-matrix least-squares on $F^2$ models within the *SHELXTL* package.[27] Data acquisition was obtained *via* Bruker *SMART* software with the corrections on Lorentz and polarization effect done with the *SAINT* program. Numerical absorption corrections due to high concentration of Pt were accomplished with *XPREP*.[28,29]

---

[†]Certain commercial equipment, instruments, or materials are identified in this document. Such identification does not imply recommendation or endorsement by the National Institute of Standards and Technology, nor does it imply that the products identified are necessarily the best available for the purpose.

**Powder Neutron Diffraction (PND):** The powder sample was loaded into a cylindrical vanadium can and sealed with an indium O-ring inside a He-filled glovebox equipped with oxygen and water sensors. The sample was then mounted on a bottom-loading closed-circuit refrigerator. PND patterns were collected at the National Institute for Standards and Technology Center for Neutron Diffraction high-resolution powder diffractometer BT-1. The data were collected using a Ge(311) monochromator (with in-pile columniation of 60'), which produced a neutron wavelength of λ = 2.0772 Å. Patterns were collected at 295.0(1) K, 150.0(1) K, and 9.0(1) K for 6 h at each temperature.

**Scanning Electron Microscope (SEM):** Crystal images and chemical compositions were measured and analyzed using a high vacuum scanning electron microscope (SEM) (JSM-6610 LV). Samples were placed on carbon tape prior to loading into the SEM chamber and were examined at 20 kV with an exposure time of 100 s.

**Physical Properties Measurements:** Physical properties measurements were conducted on a Quantum Design Dynacool Physical Property Measurement System (PPMS) with and without applied fields. Resistivity and magnetic properties were measured from 1.8 K to 350 K while heat capacity data were collected between 1.8 K and 225 K. The magnetic susceptibility is defined as χ = M/H where M is the magnetization in units of emu ($10^{-3}$ Am$^2$), and H is the applied magnetic field. A standard relaxation calorimetry method was used to measure heat capacity and the data were collected in zero magnetic field between 1.8 K and 225 K using N-type grease. All the measurements were performed on manually picked single crystal samples of MnPt$_5$P.

**Electronic Structure Calculations:** The band structure and density of states (DOS) of MnPt$_5$P were calculated using the WIEN2k program, which has the full-potential linearized augmented plane wave method (FP-LAPW) with local orbitals implemented.[30,31] The electron exchange-correlation potential parameterized by Perdew *et al*. was used to treat the electron correlation within the generalized gradient approximation.[32] The conjugate gradient algorithm was applied, and the cutoff energy was set at 500 eV. Reciprocal space integrations were completed over an 7×7×4 Monkhorst-Pack *k*-points mesh for the non-magnetic calculation and 8×8×3 for the magnetic calculation.[33] With these settings, the calculated total energy converged to less than 0.1 meV per atom. The spin-orbit coupling (SOC) effects were only applied for Pt atoms. The structural lattice parameters obtained from SC-XRD are used for both calculations for non-magnetic calculation while for magnetic calculation, the magnetic structure obtained from PND was utilized.

**Molecular Orbital (MO) Calculation:** Semi-empirical extended-Hückel-tight-binding (EHTB) methods and CAESAR packages are employed in calculating molecular orbitals of $MnPt_5P$ and $MnPt_5As$.[34] The basis sets for Mn are: *s*: Hii = -9.7500 eV, $\zeta_1$ = 0.9700, coefficient1 = 1.0000; *p*: Hii = -5.8900 eV, $\zeta_1$ = 0.9700, coefficient1 = 1.0000; *d*: Hii = -11.6700 eV, $\zeta_1$ = 5.1500, coefficient1 = 0.5139, $\zeta_2$ = 1.7000, coefficient2 = 0.6929. For Pt: *s*: Hii = -9.0770 eV, $\zeta_1$ = 2.5540, coefficient1 = 1.0000; *p*: Hii = -5.4750 eV, $\zeta_1$ = 2.5540, coefficient1 = 1.0000; *d*: Hii = -12.5900 eV, $\zeta_1$ = 6.0130, coefficient1 = 0.6334, $\zeta_2$ = 2.6960, coefficient2 = 0.5513. For P: *s*: Hii = -18.60 eV, $\zeta_1$ = 1.750, coefficient1 = 1.0000; *p*: Hii = -14.00 eV, $\zeta_1$ = 1.300, coefficient1 = 1.0000. For As: *s*: Hii = -16.2200 eV, $\zeta_1$ = 2.2300, coefficient1 = 1.0000; *p*: Hii = -12.1600 eV, $\zeta_1$ = 1.8900, coefficient1 = 1.0000.

## Results and Discussion

**Crystal Structure and Phase Determination of MnPt$_5$P:** The crystal structure of MnPt$_5$P was determined to be similar to MnPt$_5$As.[25] MnPt$_5$P crystallizes in a tetragonal unit cell with the space group of *P*4/*mmm*. Crystallographic data including refinement results, atomic coordinates, and isotropic displacement parameters are listed in Tables 1 and 2. As shown in Figure 1*a*, the crystal structure of MnPt$_5$P is analogous to MnPt$_5$As where the Mn@Pt$_{12}$ polyhedral layers are separated by pnictogen layers (P/As). The character of the layered structural feature can be observed in the SEM image in Figure S1. Both Mn and P atoms adopt into one equivalent site, respectively, while two atomic sites are occupied by Pt atoms, marked as Pt1 and Pt2. The Mn-Pt distance within the *ab*-plane is 2.751 (1) Å in MnPt$_5$P and 2.780 (1) Å in MnPt$_5$As, and the Mn–Pt–Mn angle is constrained by symmetry to be 90° in both materials. Meanwhile, the Mn-Pt distance along the *c*-axis is 2.803 (1) Å in MnPt$_5$P and 2.791 (1) Å in MnPt$_5$As. Moreover, the binary phase MnPt$_3$ consists of similar Mn@Pt$_{12}$ polyhedra with identical Mn-Pt distances of 2.758(1) Å and Mn–Pt–Mn bond angles of 90°. Interestingly, the shortest Mn-Mn distances for MnPt$_5$P, MnPt$_5$As and MnPt$_3$ are 3.891(1) Å, 3.931(1) Å, and 3.900(1) Å, respectively. Meanwhile, the Mn-Mn interatomic distances are 6.921(3) Å and 7.092(2) Å in MnPt$_5$P and MnPt$_5$As between Mn layers. MnPt$_5$As and MnPt$_3$ were reported to be ferromagnetically ordered with T$_c$ ~ 301 K and 390 K, respectively. However, MnPt$_5$P shows no ferromagnetic ordering at room temperature but an antiferromagnetic ordering at T$_N$ ~ 188 K. Details will be discussed in the following paragraphs.

Powder X-ray diffraction (PXRD) was carried out on the polycrystalline MnPt$_5$P. The PXRD pattern and Rietveld fits are illustrated in Figure 1*b*. Two additional phases, MnPt$_3$ and Pt, were included into the refinement as impurities resulting in no unindexed reflections. The goodness-of-fit parameters, $R_p$, $R_{wp}$ and $\chi^2$, were determined to be 8.03%, 11.0% and 2.39, respectively, which indicates a reasonable pattern fitting. The weight percentage of two magnetic phases, MnPt$_5$P and MnPt$_3$, appeared to be 97(1) wt% and 1.01(2) wt%.

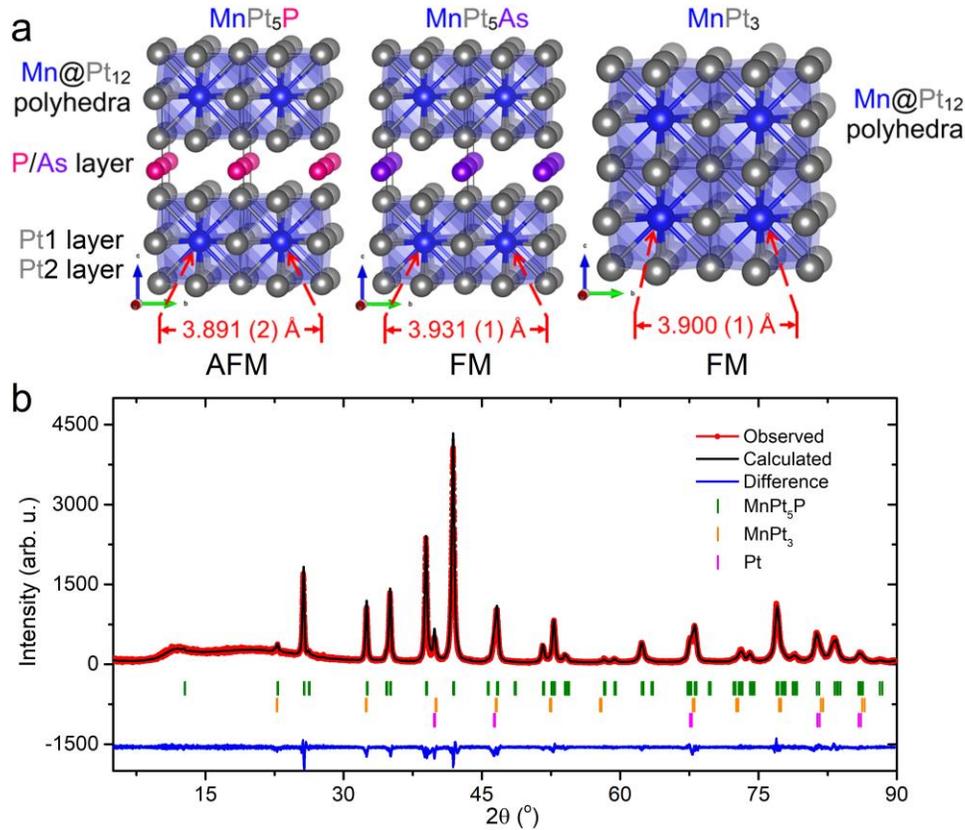

**Figure 1** *a*. Crystal structures of MnPt$_5$P, MnPt$_5$As, and MnPt$_3$ where blue, grey, red and purple spheres represent Mn, Pt, P, and As atoms, respectively. (AFM: antiferromagnetic; FM: ferromagnetic.) *b*. Powder XRD pattern of MnPt$_5$P refined by the Rietveld method. Red line with circle, black line, and blue line are the observed, calculated patterns and residual intensities, respectively. Green, orange and pink vertical ticks indicate the Bragg peaks' positions for MnPt$_5$P, MnPt$_3$ and Pt, respectively.

**Table 1.** Single crystal structure refinement for MnPt$_5$P at 296(2) K.

| Refined Formula | MnPt$_5$P |
| --- | --- |
| molar density (g/mol) | 1061.36 |
| Space group; Z | $P\,4/mmm$; 1 |
| $a$(Å) | 3.891 (2) |
| $c$(Å) | 6.921 (3) |
| V (Å$^3$) | 104.8 (1) |
| θ range (º) | 2.943-33.023 |
| No. reflections; $R_{int}$ | 1109; 0.0319 |
| No. independent reflections | 154 |
| No. parameters | 12 |
| $R_1$: $\omega R_2$ ($I>2\delta(I)$) | 0.0173; 0.0450 |
| Goodness of fit | 1.317 |
| Diffraction peak and hole (e$^-$/ Å$^3$) | 2.845; -3.625 |

**Table 2.** Atomic coordinates and equivalent isotropic displacement parameters for MnPt$_5$P at 296(2) K. ($U_{eq}$ is defined as one-third of the trace of the orthogonalized $U_{ij}$ tensor (Å$^2$)). Values in parentheses indicate one standard deviation.

| Atom | Wyckoff. | Occ. | x | y | z | $U_{eq}$ |
|------|----------|------|---|---|---|----------|
| Pt1  | 1a       | 1    | 0 | 0 | 0 | 0.0032(2) |
| Pt2  | 4i       | 1    | 0 | ½ | 0.2916(1) | 0.0032(2) |
| Mn3  | 1c       | 1    | ½ | ½ | 0 | 0.0034(6) |
| P4   | 1b       | 1    | 0 | 0 | ½ | 0.004(1) |

**Magnetic Properties of MnPt$_5$P Crystals:** The magnetic properties of MnPt$_5$P were measured on select single crystals. The single crystals were pre-aligned using the single crystal X-ray diffractometer before loading. The measurements were performed on the same crystallites with two distinct orientations. One is perpendicular to the *c*-axis of the crystal (B⊥*c*), as shown in Figure 2*a*, *c* & *e*; and the other one is parallel to the *c*-axis (B//*c*), shown in Figure 2 *b*, *d* & *f*.

The temperature-dependence of magnetic susceptibility along two different directions as well as their inversed susceptibility are illustrated in Figure 2*a* & *b*. The measurements were conducted under an applied magnetic field of 1000 Oe ((1000/4π) A/m) from 1.8 K to 350 K. A large difference in susceptibility values could be easily found along two directions (the susceptibility ~ 5 emu/Oe/mol (1 emu/Oe/mol = 4π 10$^{-6}$ m$^3$/mol) for B⊥*c*; but ~1.2 emu/Oe/mol for B//*c* at 1.8 K). This indicates strong magnetic anisotropy. Meanwhile, a sharp antiferromagnetic ordering transition peak could be observed in both sweep directions at a Néel temperature ($T_N$) ~ 188 K. Below the transition temperature, the magnetic susceptibility started dropping with decreasing temperature and reached the minimum at ~147 K. Below ~147 K, the magnetic susceptibility increased and achieved a plateau below 50 K under the Field-Cooling (FC) measurement. In the Zero-Field-Cooling (ZFC) mode, the magnetic susceptibility showed similar trends, except that the magnetic susceptibility decreased below 50 K. Above the Néel temperature, the inverse susceptibility did not show a linear behavior due to the slight ferromagnetic MnPt$_3$ impurity which orders at ~390 K.

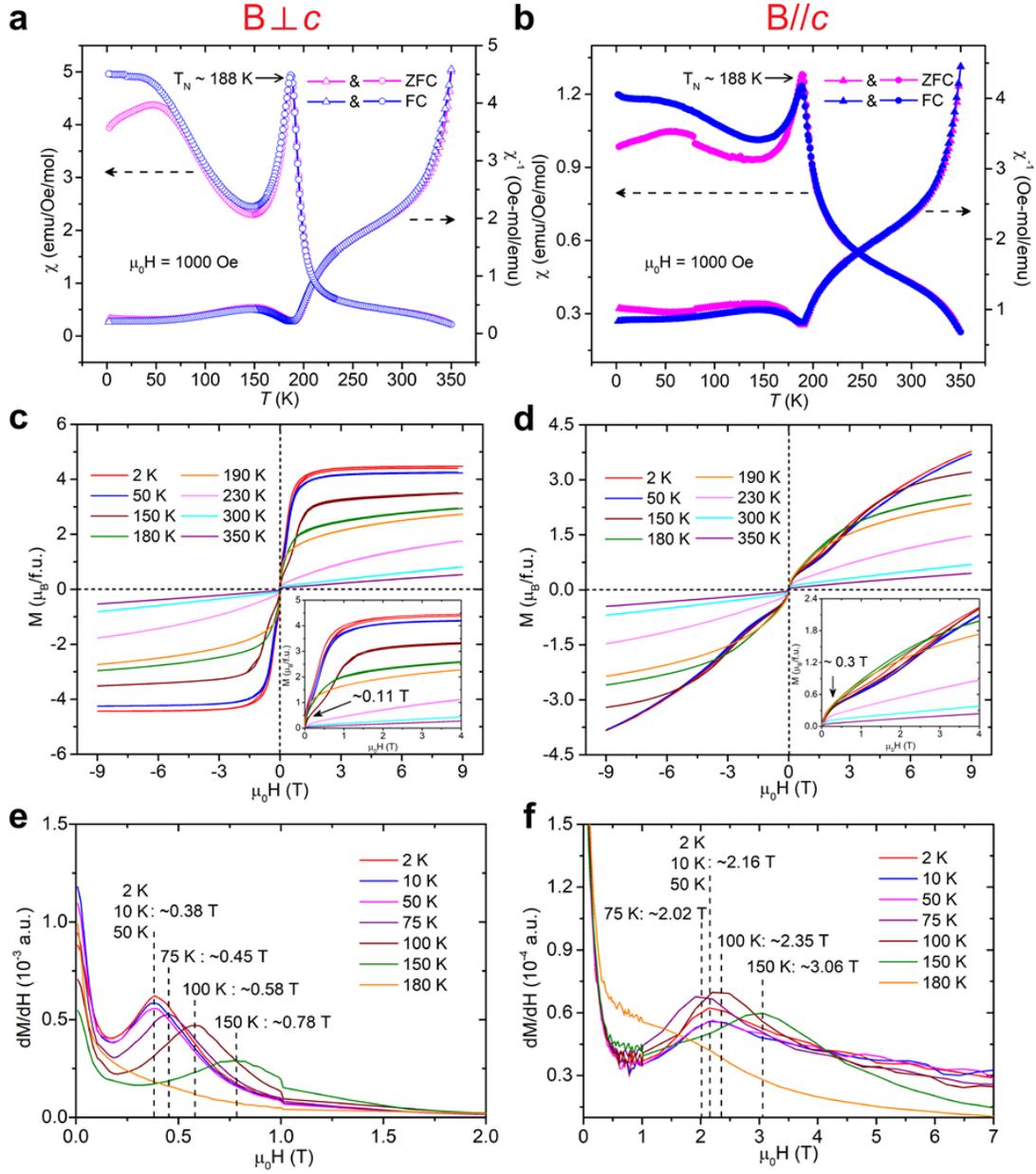

**Figure 2.** Temperature-dependence of magnetic susceptibility for MnPt$_5$P under an applied magnetic field of 1000 Oe when field is ***a.*** perpendicular to the *c*-axis (B⊥*c*); ***b.*** parallel to the *c*-axis (B//*c*). Field-dependence of magnetization for MnPt$_5$P under various temperatures when field is ***c.*** perpendicular to the *c*-axis (B⊥*c*); ***d.*** parallel to the *c*-axis (B//*c*). ***e.*** & ***f.*** present dM/dH *vs* applied field indicating the spin-flop field. 1 Oe = (1000/4π) A/m.

Figure 2*c* & *d* present the hysteresis loops of MnPt$_5$P under various temperatures. A small coercive field (< 200 Oe) was detected, an indication of soft magnetic behaviors in MnPt$_5$P. Figure 2*c* showed a ferromagnetic-like loop. A clear magnetic anisotropy could be seen based on the distinct saturated field for two field orientations. At 2 K, the saturated moment for B⊥$c$ is ~4.49 $\mu_B$/f.u. while no saturation was achieved for the other direction. This indicates that in MnPt$_5$P, the easy axis lies within the *ab*-plane. Meanwhile, a metamagnetic phase transition can be seen by referring to the upturn of magnetization with increasing magnetic field at low-field regions for both directions below T$_N$, as shown in the insets of Figure 2*c* & *d*. In Figure 2*e* & *f*, $d$M/$d$H with respect to the applied magnetic field showed peaks for this metamagnetic phase transition, which could be a spin-flop transition. During the transition, the spin-flop field, H$_{SF}$, moved towards higher fields with increasing temperature. The value of H$_{SF}$ also exhibited strong magnetic anisotropy referring to the large difference between H$_{SF}$s for the two directions. Figure 3 presents the magnetic phase diagram of MnPt$_5$P. The data were extracted from Figure 2*e* & *f*. Above T$_N$ ~ 188 K, MnPt$_5$P is paramagnetic while below ~150 K, a spin-flop transition emerged in both crystallographic orientations. Notably, after the spin-flop transition (as shown by the post-SF region in Figure S2), MnPt$_5$P shows ferromagnetic ordering with a saturated moment when B⊥$c$ as discussed previously. However, MnPt$_5$P did not exhibit saturated magnetization below 9 T when B//$c$.

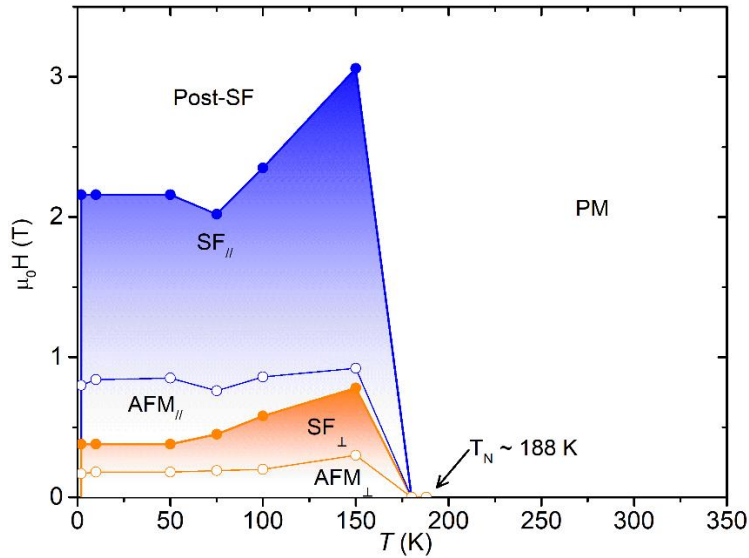

**Figure 3.** Magnetic phase diagram of MnPt$_5$P. PM: paramagnetic state; AFM: antiferromagnetic state; SF: spin-flop. The open and solid circles were extracted from Figure 2.*e* & *f* which represent the beginning and the end of spin-flop transition, respectively. //: external magnetic field is applied parallelly to *c* axis; ⊥: external magnetic field is applied perpendicularly to *c* axis.

**Magnetic Structures of MnPt₅P:** To determine the magnetic structure of MnPt$_5$P, the powder neutron diffraction patterns at 295 K, 150 K, and 9 K were analyzed using the Topas Academic,[35] EXPGUI/GSAS,[36,37] and FullProf suite[38] software packages. An initial Pawley fit of the pattern collected at 295 K again revealed Pt and MnPt$_3$ as minority impurity phases.[39,40] A subsequent thorough Rietveld refinement of the MnPt$_5$P phase yielded the nuclear crystal structure at 295 K.[41] During this analysis, the total scattering arising from the ferromagnetic MnPt$_3$ impurity was treated as a single phase using a Pawley phase, while the Pt phase was treated using a full Rietveld refinement.

At 150 K, six additional Bragg peaks appeared in the diffraction pattern. Given that Pt is diamagnetic, MnPt$_3$ is ferromagnetic, and that the antiferromagnetic ordering temperature for MnPt$_5$P is $T_N \approx 188$ K, the new Bragg peaks were assigned as magnetic Bragg peaks arising from long-range antiferromagnetic ordering in MnPt$_5$P. The new Bragg peaks were indexed to a doubling of the unit cell in the *c*-axis direction for the MnPt$_5$P phase. At 150 K, the resulting Miller indices for the six new Bragg peaks are (001), (003), (101), (103), (111), and (113) at Q ≈ 0.46 Å$^{-1}$, 1.37 Å$^{-1}$, 1.68 Å$^{-1}$, 2.12 Å$^{-1}$, 2.33 Å$^{-1}$, and 2.65 Å$^{-1}$, respectively (Figure 3). Each pattern is plotted individually (Figures S2−S4). and an enhanced view of the magnetic Bragg peaks in the 9 K pattern with the corresponding refinement curve, which illustrates the quality of the refinement, is shown in the supporting document (Figure S5).

The *k*-search functionality in FullProf confirmed a commensurate magnetic propagation vector of $\vec{k}$ = (00½). Representational analysis of the $\vec{k}$(00½) propagation vector in the parent *P4/mmm* space group using SARA*h*[33] in conjunction with the FullProf suite led to three unique irreducible representations with basis vectors, which are summarized in Table 3.

**Table 3.** Irreducible represenatations (IR) and corresponding basis vectors (BV) for the magnetic Mn ion at the (½, ½, 0) fractional coordinates and associated real magnetic components in the *a*-, *b*-, and *c*-axis directions for the $\vec{k}$(00½) propagation vector in the *P4/mmm* space group.

| IR | BV | Basis Vector Components | | |
|---|---|---|---|---|
| | | $m_{\parallel a}$ | $m_{\parallel b}$ | $m_{\parallel c}$ |
| Γ₃ | ψ₁ | 0 | 0 | 16 |
| Γ₉ | ψ₂ | 8 | 0 | 0 |
| Γ₉ | ψ₃ | 0 | −8 | 0 |

Each irreproducible representation accounts for magnetic scattering in a single unit cell direction in real space. Given the non-zero intensity of the (111) magnetic Bragg peak, all three irreducible representations were required to accurately model the antiferromagnetic phase for MnPt$_5$P at 150 K. The basis vector mixing coefficients were refined freely during the Rietveld refinement fitting for the 150 K pattern. The same analysis was conducted for the neutron powder diffraction pattern collected at 9 K. At 9 K, the pattern shows an increase in intensity of the six magnetic Bragg peaks arising from MnPt$_5$P, corresponding to an increase in the magnitude of the magnetic moments. The analysis of the nuclear and magnetic phases resulted in the magnetic structure for MnPt$_5$P at 150 K and 9 K as shown in Figure 5. These two magnetic structures agree qualitatively. At both temperatures, the material is a canted A-type antiferromagnetic with the spins nearly aligned along the (110) crystallographic plane in real space, with canting in the *c*-axis direction. The resulting magnetic moments for the Mn ions are summarized in Table 4.

The A-type antiferromagnetism in MnPt$_5$P is distinct from the ferromagnetism exhibited by the isostructural MnPt$_5$As and the structurally similar MnPt$_3$ compounds. The magnetic measurements show that, by chemical substitution with P, the interlayer Mn–Mn FM interaction is replaced by an interlayer AFM exchange coupling interaction. A simple Goodenough-Kanamori type analysis of the MnPt$_5$P crystal structure suggests that the magnitude of the interlayer FM superexchange interaction should be enhanced in MnPt$_5$P compared to MnPt$_5$As because the bond angles along the Mn–Pt–Pnictogen–Pt–Mn pathway remain relatively constant while the bond lengths decrease significantly with P substitution assuming that the nature of the Pt-Pnictogen bonding remains constant. Therefore, we hypothesized that a change in the nature of the Pt–Pnictogen bonding effectively drives the magnitude of the interlayer FM superexchange interaction to zero, and a through-space AFM interaction becomes the dominant interlayer magnetic exchange interaction. To test this hypothesis, we performed DFT calculations to investigate the nature of the metal-pnictogen bonding near the Fermi surface.

**Table 4.** The projection on each crystal axis from magnetic moments on Mn in MnPt$_5$P. Values in parentheses indicate one standard deviation.

| Translation | Crystal axis | Moments at 9 K/$\mu_B$ | Moments at 150 K/$\mu_B$ |
|---|---|---|---|
| (0, 0, 0) | a | 1.997(1) | 1.394(1) |
|  | b | −2.129(1) | −1.740(1) |
|  | c | 0.887(1) | 1.108(1) |
| (0, 0, 1) | a | −1.997(1) | −1.394(1) |

|  | b | 2.129(1) | 1.740(1) |
|  | c | −0.887(1) | −1.108(1) |
| **Total moment** |  | 3.051 | 2.490 |

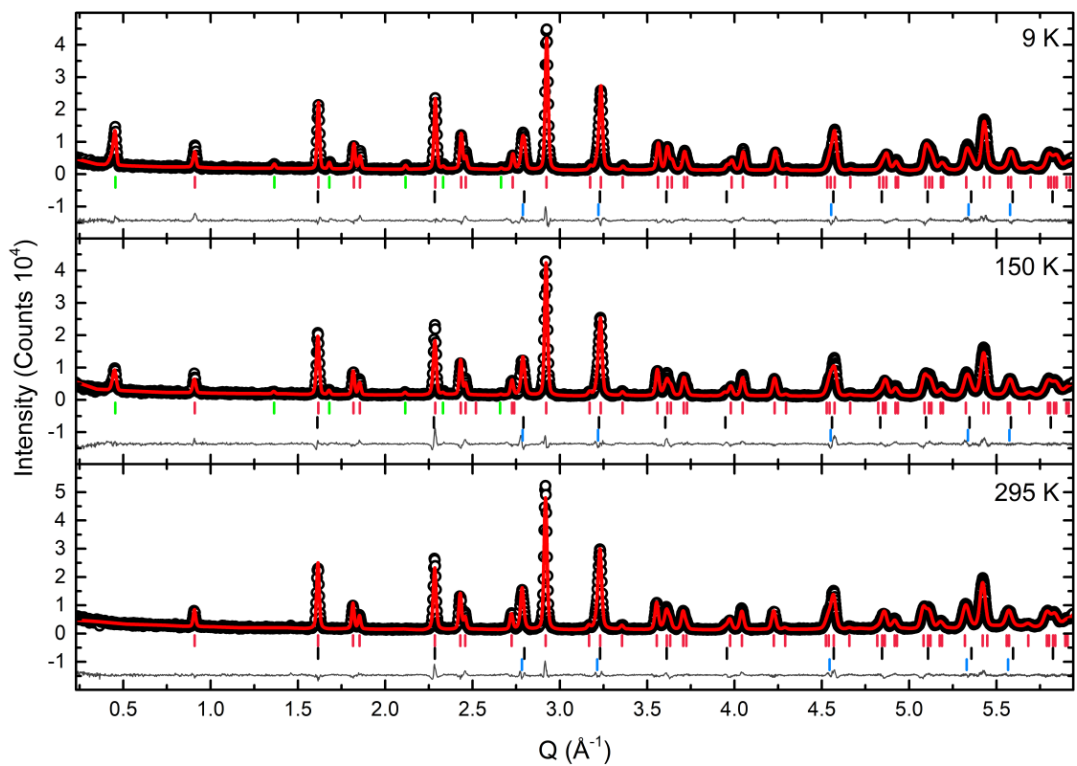

**Figure 4.** Fitted powder neutron diffraction patterns of MnPt$_5$P at *a.* 9 K; *b.* 150 K; *c.* 295 K. The black circle, red line and grey line stand for the observed pattern, calculated pattern, and residual intensities. Red, blue, and black vertical ticks represent the Bragg peak positions for MnPt$_5$P, Pt, and MnPt$_3$. Green vertical tick marks denote the magnetic Bragg peak positions for MnPt$_5$P.

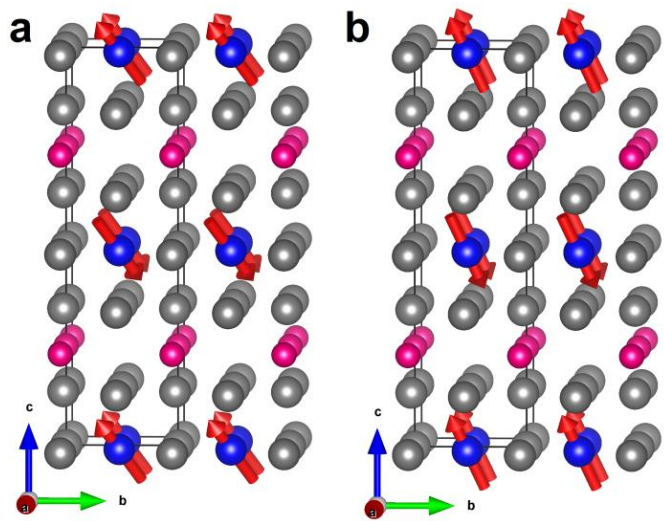

**Figure 5.** Magnetic structures of MnPt$_5$P at *a.* 9 K; *b.* 150 K.

**Resistivity and Heat Capacity Measurements:** Resistivity measurement was carried out between 2 K and 350 K without an applied magnetic field on the single crystal measured for magnetic properties. Four-probe method was employed to make the electrical contacts with Ag paint. The temperature-dependence of resistivity shown in Figure 6a demonstrates metallic behavior with the small resistivity. The high RRR value ($\rho_{300K}/\rho_{2K}$ ~23.6) indicates the high quality and lack of defects in the crystal. A sudden drop in resistivity could be found around $T_N$ ~ 188 K, which is consistent with the antiferromagnetic ordering transition of MnPt$_5$P. $T^1$ and $T^3$ behaviors were fitted for high-temperature (260 K - 350 K) and low-temperature (2 K - 30 K) regions, respectively, by using the following equation: $\rho(T) = \rho_0 + AT^n$ where $\rho_0$ is the residual resistivity due to defect scattering, A is a constant, and *n* is an integer determined by the interaction pattern. The results implied that above $T_N$, the resistivity of MnPt$_5$P was mainly generated by electron-phonon collision while at low temperature, the *s-d* electron scattering took over.

Heat capacity measurements were conducted between 1.8 K and 225 K without an applied magnetic field on the same crystal measured for resistivity and magnetism. As can be seen in Figure 6b, with decreasing temperature, an abrupt and large upturn could be found below 195 K and the maximum was reached at ~188 K, where antiferromagnetic ordering occurs. Other than that, no more phase transition peaks could be found in the lower temperature region. Due to the lack of determination of the phonon mode contributions, the magnetic entropy change could not be calculated. Magnetism, resistivity, and heat capacity results are in good agreement with each other.

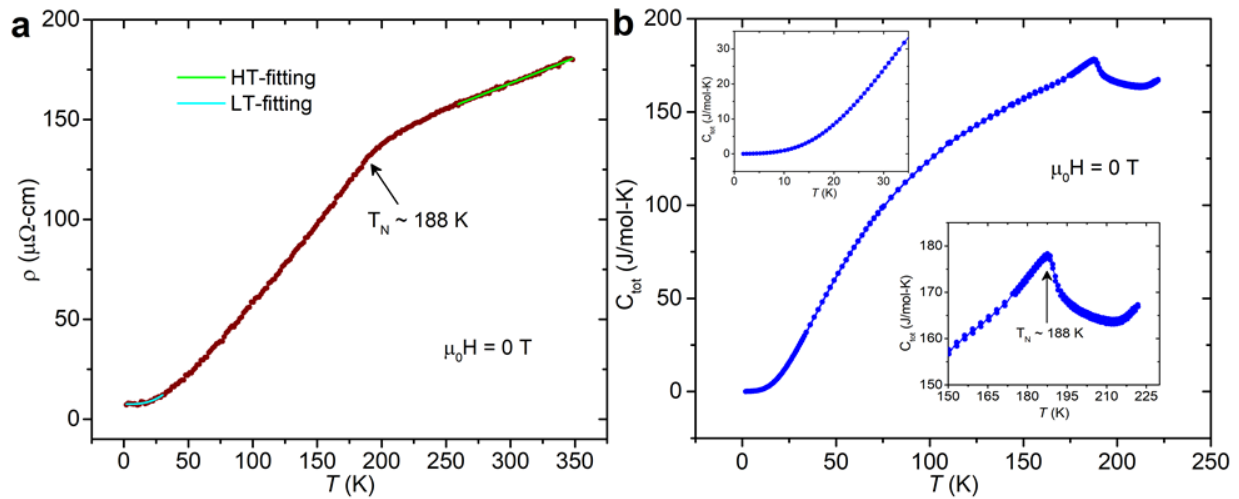

**Figure 6a.** Temperature-dependence of resistivity of MnPt$_5$P between 2 K and 350 K. Green and cyan lines represent the fitting lines for high-temperature (HT) and low-temperature (LT) regions. **b.** Heat capacity curve of MnPt$_5$P from 1.8 K to 225 K.

**Electronic Structures and Molecular Orbital Diagram:** The electronic structure of MnPt$_5$P was calculated including spin-orbit coupling (SOC) effects and spin-polarization (SP). The electronic structure calculation with SP was performed with the magnetic unit cell shown in Figure 7*c* labelled with atomic sites, i.e., with a doubled unit cell of the crystal structure, and the Brillouin Zone (BZ) appeared in Figure 7*d*.

Figure 7*a* displays the band structure of MnPt$_5$P. The dense bands observed around the Fermi level indicate the metallic properties of MnPt$_5$P, which is consistent with the resistivity measurements. One can see that with consideration of SOC effect, the bands near the Fermi level ($E_F$) split, i.e., at the Γ point. By taking SP into account, the bands were found to be denser below and above $E_F$ due to a doubling of the unit cell. However, the bands near $E_F$ were fewer and more dispersive, that is, fewer "flat bands". The projection from crucial orbitals of atoms on distinct atomic sites was listed in Figure S7. This analysis revealed that Mn-*d* and Pt-*d* orbitals were dominant near $E_F$. However, the most significant contribution to the DOS from the P atoms came from the P-*p* orbital, but this was not as critical as Mn-*d* and Pt-*d* orbitals near $E_F$. With inclusion of SOC effect, the band intensity was dramatically decreased. While SOC and SP were both considered, the Mn1 site dominates near $E_F$ with the Mn2 site more effective above the Fermi level. The density of states (DOS) corresponding to the calculated band structures are presented in Figure 7*b*. The DOS at $E_F$ is reduced when SOC was considered and dropped dramatically when both SOC and SP were included in the calculations. In Figure 7*b* (Right), a pseudogap appears at ~0.4 eV above $E_F$ and a van Hove singularity emerges ~ −0.25 eV. The van Hove singularity is alleviated when SP effects are included in the calculation. Moreover, as can be seen in the inset of Figure 7*b* (Right), the contribution from Mn1 and all Pt sites were dominant at the Fermi level. The low density of states comparing with other calculations without consideration of SP is due to the antiferromagnetic ordering of Mn atoms.

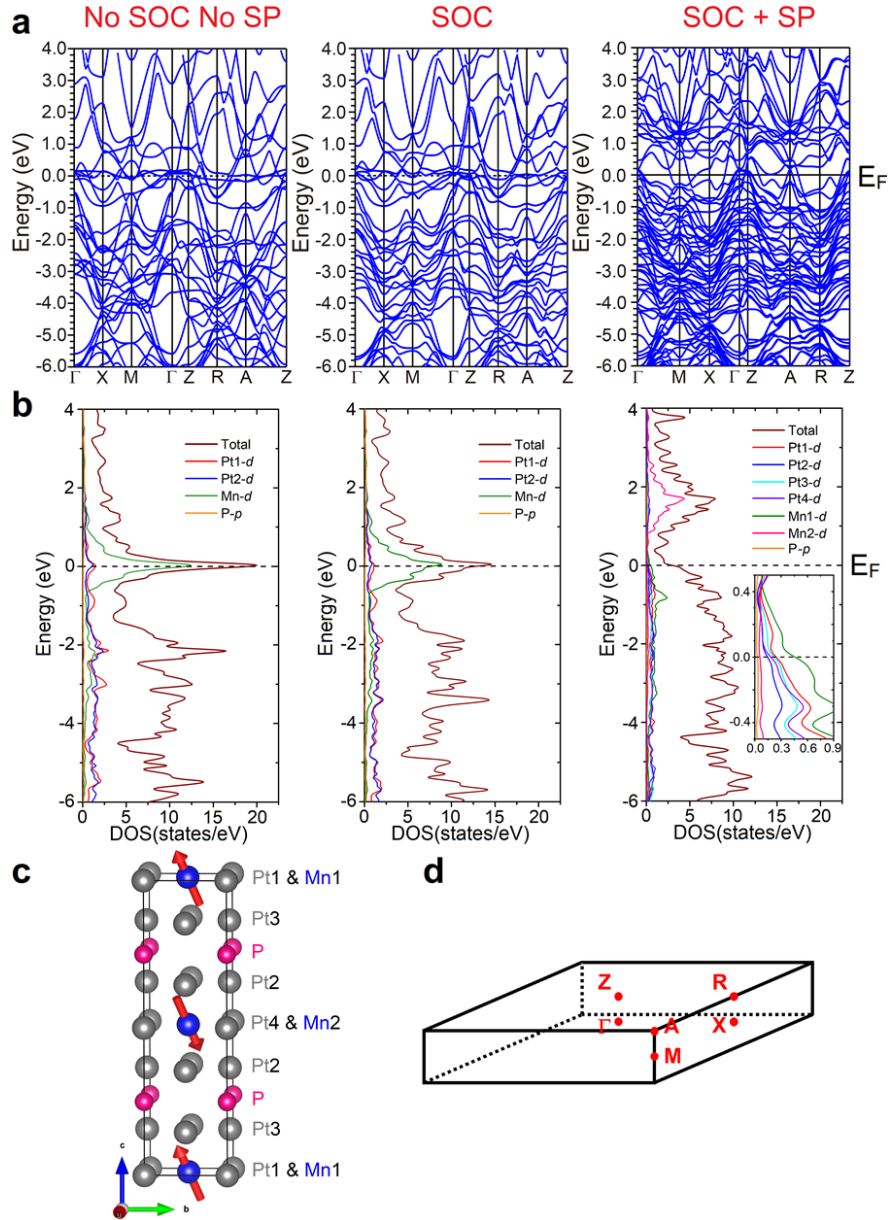

**Figure 7** *a.* Band structures calculated for MnPt$_5$P with/without consideration of SOC effect and SP. *b.* Density of states corresponding to *a*. *c.* Magnetic structure of MnPt$_5$P at 9 K with atomic sites labelled. *d.* Brillouin zone utilized for calculation with SP.

**Chemical bonding in MnPt$_5$P and MnPt$_5$As:** To estimate the chemical bond influence on atomic interactions and magnetic behavior in MnPt$_5$P and MnPt$_5$As, molecular orbital diagrams were generated based on the semi-empirical extended-Hückel-tight-binding (EHTB) methods using the CAESAR packages, shown in Figure 8. All the molecular orbitals exhibited antibonding features in both MnPt$_5$P and MnPt$_5$As. The lowest unoccupied molecular orbitals (LUMOs) for both compounds appeared the same where Mn, Pt1 and Pt2 atoms show $d_{x^2-y^2}$, $d_{x^2-y^2}$ and $d_{z^2}$ character. No contribution could be found from the pnictogen atomic sites. However, significant differences are observed when it came to the highest occupied molecular orbitals (HOMOs). MnPt$_5$P was found to adopt one HOMO where $d_{z^2}$ orbitals of Mn and Pt1 were dominant while Pt2 presented $d_{xz}$ character and no contribution was found for the P atom. On the other hand, in MnPt$_5$As, two degenerate HOMOs appeared (Figure 8) and showed similar components. Mn and Pt1 in the HOMOs of MnPt$_5$As are dominated by $d_{xz}$ and $d_{yz}$ orbitals while $d_{x^2-y^2}$ and $d_{z^2}$ were important at the Pt2 site. Interestingly, even though the P atoms in MnPt$_5$P did not show any major contribution in the HOMO, the As atoms in MnPt$_5$As showed strong $p$ character. Thus, we speculate that the lack of available $p$ character in P may result in shorter bond distance, reduce possible orbital overlap, and differentiate magnetic superexchange mechanisms from MnPt$_5$As. This allows an AFM interlayer through space magnetic coupling, presumably present in both pnictogen congeners, to dominate the interlayer magnetic exchange interaction in MnPt$_5$P.

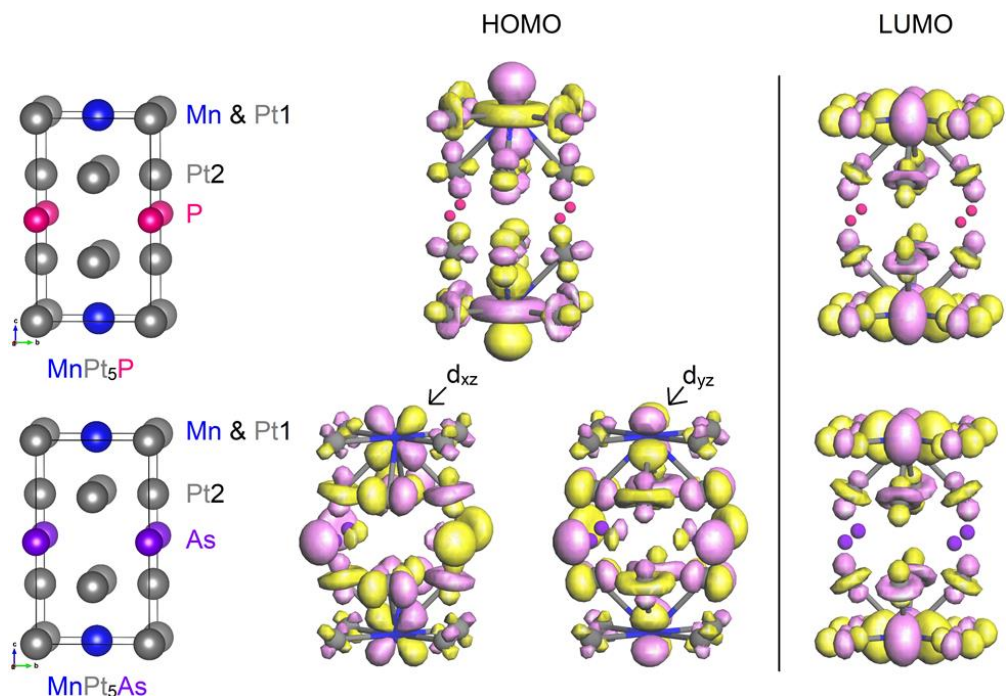

**Figure 8.** The highest occupied molecular orbitals (HOMOs) and lowest unoccupied molecular orbitals (LUMOs) of MnPt$_5$P and MnPt$_5$As.

## Conclusion

A novel antiferromagnet, MnPt$_5$P, was designed and synthesized using high temperature solid-state reaction. MnPt$_5$P contains a layered 2D crystal structure and displays magnetic anisotropy. A spin-flop transition was detected when high magnetic fields were applied. The magnetic structure was determined to be an A-type antiferromagnetic by powder neutron diffraction experiments. The theoretical calculations showed the crucial significance of *d* electrons from Mn and Pt atoms on structural stability and magnetic behaviors. The comparison on molecular orbital calculation results between MnPt$_5$P and the previously reported ferromagnetic MnPt$_5$As revealed that the lack of P-*p* character in the MnPt$_5$P HOMO may explain the antiferromagnetic ordering. MnPt$_5$P and reported MnPt$_5$As provide an ideal platform to tune ferromagnetic and antiferromagnetic ordering in the family compounds with the same structure and valence electron counts either by chemical doping or by the application of extreme pressures.

## Supporting Information

Supplementary information, details of author contributions and competing interests; and statements of data are available at https://doi.org/xxx including:

SEM image of MnPt$_5$P crystal; Powder neutron diffraction patterns with Rietveld refinement fit curve; Projections from critical orbitals of different atoms on band structure.

## Acknowledgements

The work at LSU is supported by Beckman Young Investigator award and NSF-DMR-1944965. R.A.K. acknowledges research support from the U.S. Department of Energy, Office of Energy Efficiency and Renewable Energy, Fuel Cell Technologies Office, under Contract DE-AC36-08GO28308.

**For table of content only**

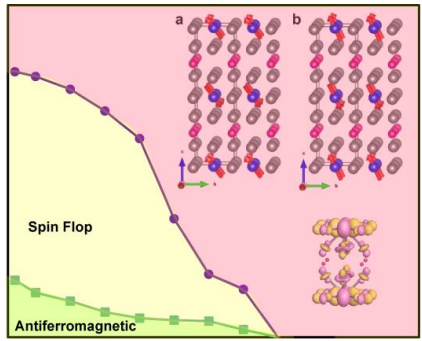

# Supplementary Information

# Chemical Bonding Governs Complex Magnetism in MnPt$_5$P

*Xin Gui,[1] Ryan A. Klein,[2,3] Craig M. Brown,[3] Weiwei Xie[1]\**

1. Department of Chemistry, Louisiana State University, Baton Rouge, LA, 70803, USA
2. Chemistry and Nanoscience Department, National Renewable Energy Laboratory, Golden, CO, 80401, USA
3. NIST Center for Neutron Research, National Institute of Standards and Technology, Gaithersburg, MD, 20899, USA

## Table of Contents

**Figure S1.** SEM image of MnPt$_5$P crystal with layered feature observed.

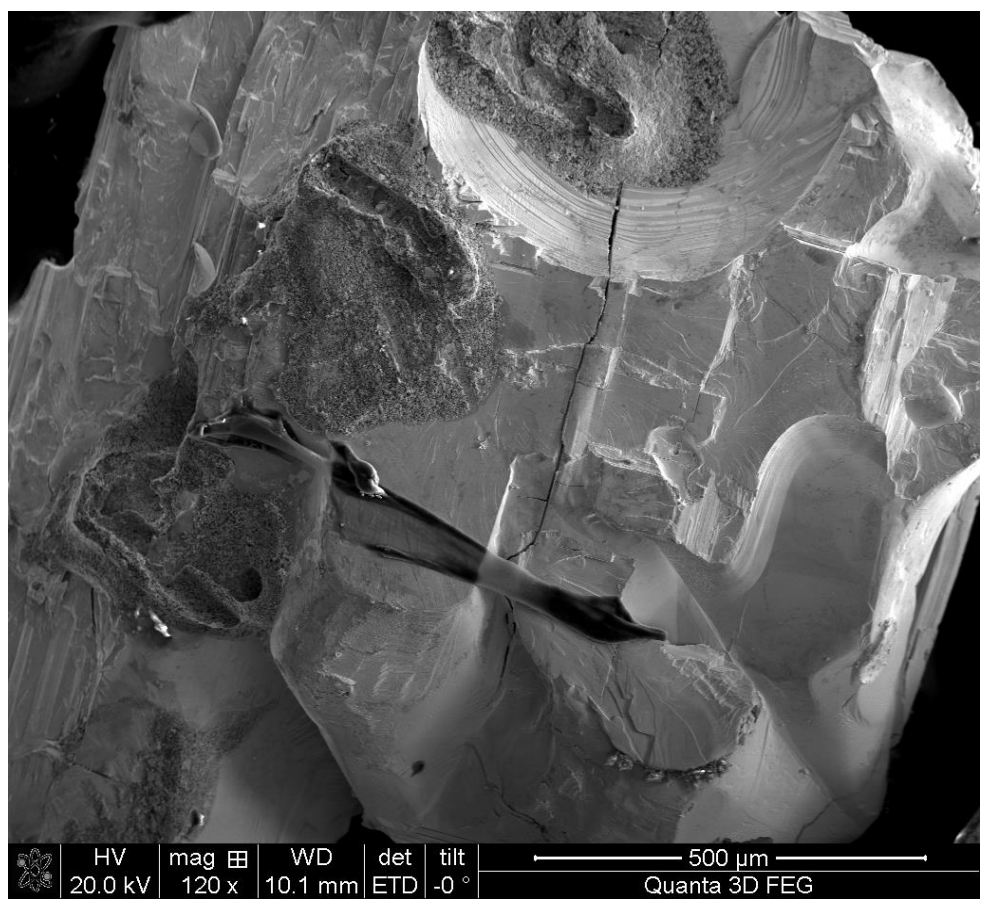

**Figure S2.** Powder neutron diffraction pattern collected at 295 K and accompanying Rietveld refinement fit curve. The black circles, red curve, and grey curve represent the data, Rietveld refinement fit curve, and the residual intensity, respectively. The vertical red tick mark, blue tick marks, and black tick marks denote the Bragg peak positions for the $MnPt_5P$, $MnPt_3$, and Pt phases, respectively. Symbols are commensurate with their error bars, which indicate one standard deviation. $R_p$ = 5.81, $R_{wp}$ = 7.22, $R_{exp}$ = 4.81, GOF = 2.26.

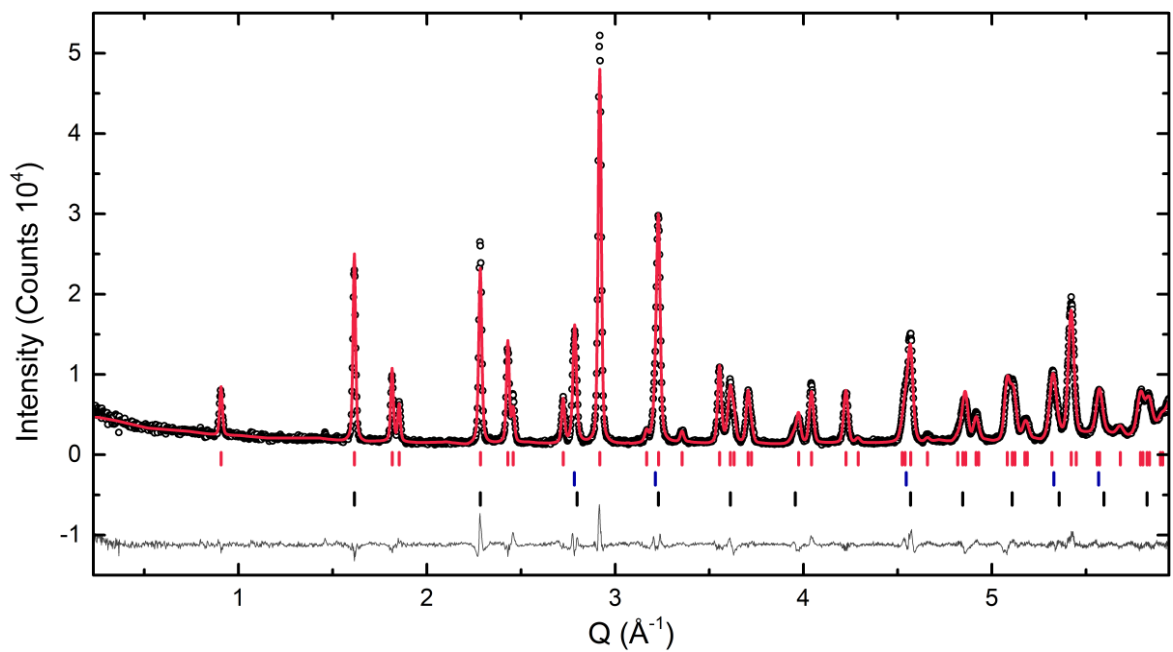

**Figure S3.** Powder neutron diffraction pattern collected at 150 K and accompanying Rietveld refinement fit curve. The black circles, red curve, and grey curve represent the data, Rietveld refinement fit curve, and the residual intensity, respectively. The vertical red tick mark, blue tick marks, and black tick marks denote the Bragg peak positions for the MnPt$_5$P, MnPt$_3$, and Pt phases, respectively. Vertical green tick marks show the magnetic Bragg peaks for MnPt$_5$P and their labels indicate the Miller indices for these reflections. Symbols are commensurate with their error bars, which indicate one standard deviation. $R_\text{p}$ = 7.08, $R_\text{wp}$ = 8.89, $R_\text{exp}$ = 5.39, GOF = 2.71, $R_\text{mag}$ = 8.99.

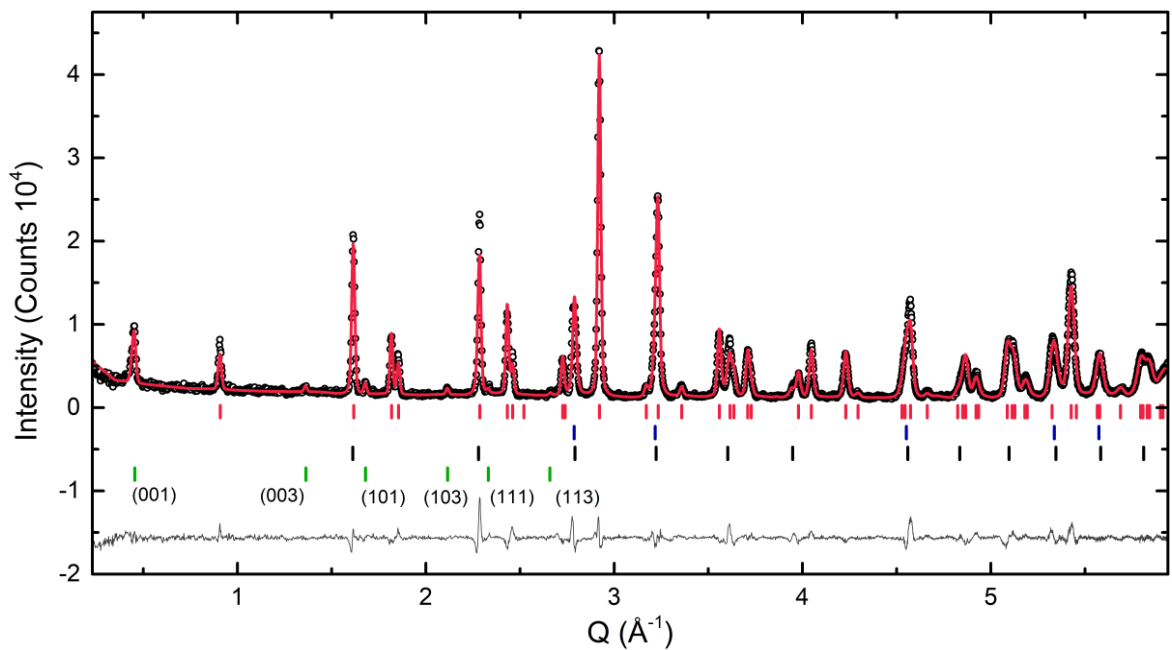

**Figure S4.** Powder neutron diffraction pattern collected at 9 K and accompanying Rietveld refinement fit curve. The black circles, red curve, and grey curve represent the data, Rietveld refinement fit curve, and the residual intensity, respectively. The vertical red tick mark, blue tick marks, and black tick marks denote the Bragg peak positions for the MnPt$_5$P, MnPt$_3$, and Pt phases, respectively. Vertical green tick marks show the magnetic Bragg peaks for MnPt$_5$P and their labels indicate the Miller indices for these reflections. Symbols are commensurate with their error bars, which indicate one standard deviation. $R_p$ = 6.06, $R_{wp}$ = 7.56, $R_{exp}$ = 5.37, GOF = 1.98, $R_{mag}$ = 5.62.

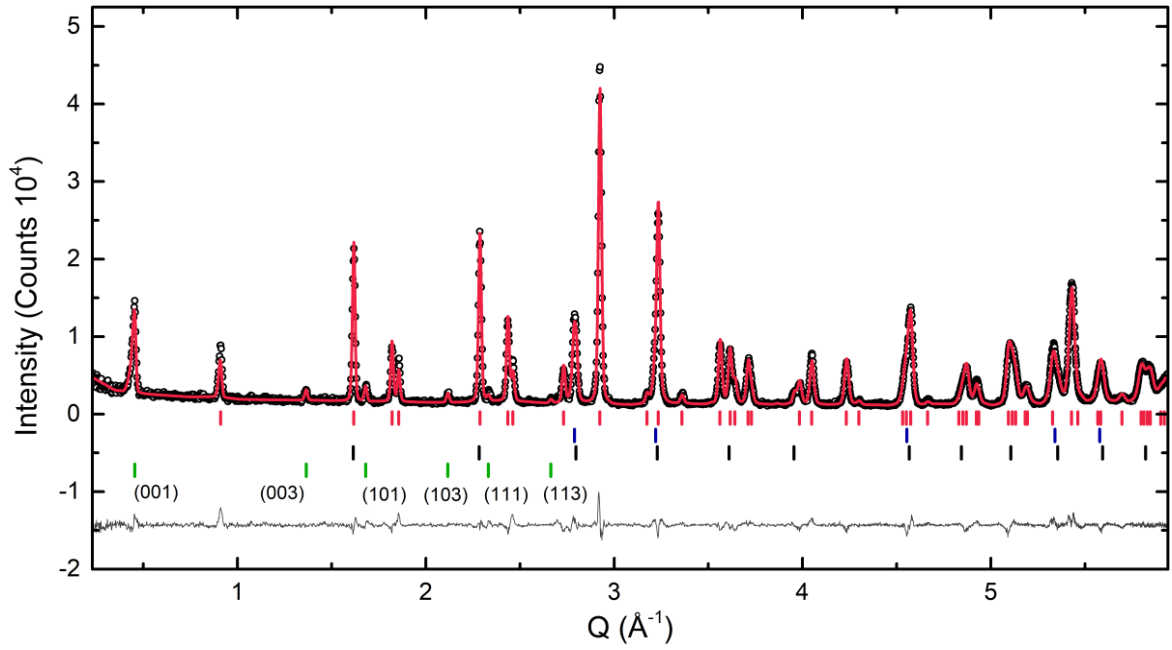

**Figure S5.** A zoomed-in view of each of the magnetic Bragg peaks in the 9 K powder neutron diffraction pattern shows the quality of the Rietveld refinement for the MnPt$_5$P magnetic phase. The black circles, red curve, and grey curve represent the data, Rietveld refinement fit curve, and the residual intensity, respectively. The vertical red tick mark, blue tick marks, and black tick marks denote the Bragg peak positions for the MnPt$_5$P, MnPt$_3$, and Pt phases, respectively. Vertical green tick marks show the magnetic Bragg peaks for MnPt$_5$P and their labels indicate the Miller indices for these reflections. Symbols are commensurate with their error bars, which indicate one standard deviation. The fit statistics are listed in the figure caption for Fig. S4.

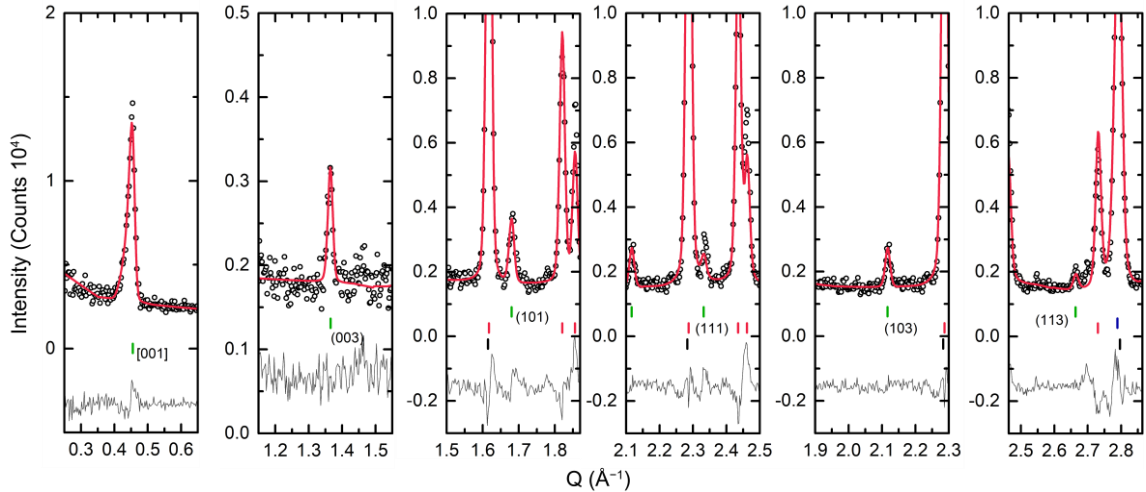

**Figure S6.** Projections from critical orbitals of different atoms on band structure.

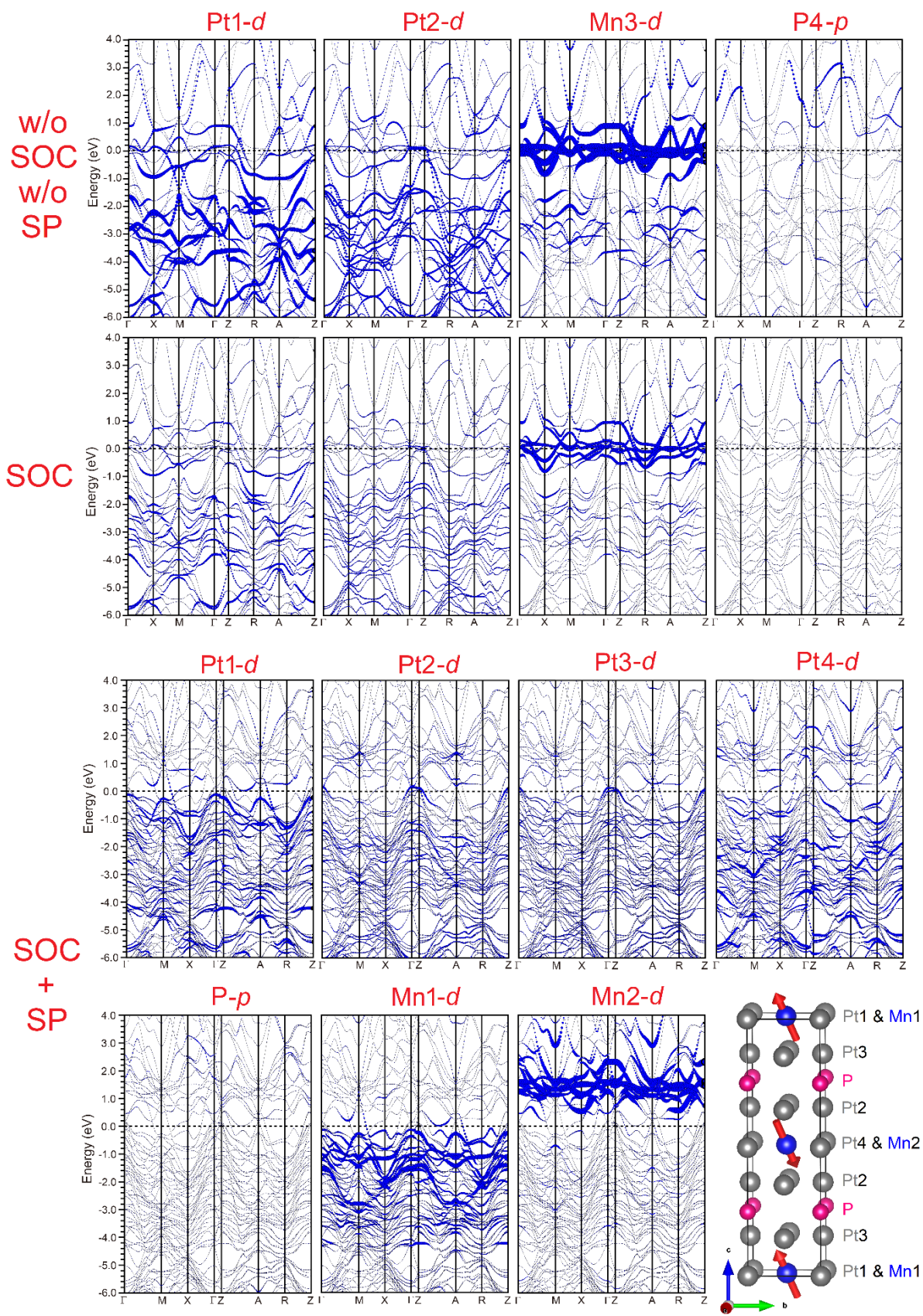

**Table S1.** Atomic coordinates and isotropic thermal parameters for MnPt$_5$P at 295 K, 150 K, and 9 K as obtained from the Rietveld refinements of the powder neutron diffraction data. Values in parentheses indicate one standard deviation.

| Temp (K) | Atom | Wyckoff. | Occ. | $x$ | $y$ | $z$ | $U_{iso}$ |
|---|---|---|---|---|---|---|---|
| 295.0(1) | Pt1 | 4$i$ | 1 | 0 | ½ | 0.2912(2) | 0.0014(4) |
|  | Pt2 | 1$a$ | 1 | 0 | 0 | 0 | 0.0023(8) |
|  | Mn3 | 1$c$ | 1 | ½ | ½ | 0 | 0.002(1) |
|  | P4 | 1$b$ | 1 | 0 | 0 | ½ | 0.006(2) |
| 150.0(1) | Pt1 | 4$i$ | 1 | 0 | ½ | 0.2907(2) | 0.0015(4) |
|  | Pt2 | 1$a$ | 1 | 0 | 0 | 0 | 0.0002(9) |
|  | Mn3 | 1$c$ | 1 | ½ | ½ | 0 | 0.002(1) |
|  | P4 | 1$b$ | 1 | 0 | 0 | ½ | 0.007(2) |
| 9.0(1) | Pt1 | 4$i$ | 1 | 0 | ½ | 0.2913(2) | 0.0024(4) |
|  | Pt2 | 1$a$ | 1 | 0 | 0 | 0 | 0.0010(7) |
|  | Mn3 | 1$c$ | 1 | ½ | ½ | 0 | 0.005(1) |
|  | P4 | 1$b$ | 1 | 0 | 0 | ½ | 0.009(1) |